\documentstyle[prd,aps,preprint]{revtex}
\begin{document}
\draft

%\advance\textwidth -1in

\title{Black-hole collisions from Brill-Lindquist initial
data: predictions of perturbation theory}

\author{Andrew M. Abrahams} \address{Department of Physics and
Astronomy, University of North Carolina, Chapel Hill, NC 27599-3255}
\author{Richard H. Price} \address{Department of Physics, University
of Utah, Salt Lake City, Utah, 84112}

\date{September 9, 1995}
\maketitle
%\tightenlines

\begin{abstract}
\widetext The Misner initial value solution for two momentarily
stationary black holes has been the focus of much numerical study. We
report here analytic results for an astrophysically similar initial
solution, that of Brill and Lindquist (BL). Results are given from
perturbation theory for initially close holes and are compared with
available numerical results. A comparison is made of the radiation
generated from the BL and the Misner initial values, and the physical
meaning is discussed.
\end{abstract}
%insert suggested PACS numbers in braces on next line
%\pacs{04.70.-s, 04.30.-w, 04.25.Dm, 04.25.Nx}

\section{Introduction}\label{sec:intro}
Black hole collisions are presently of great recent interest as one of
the ``grand challenges'' in high performance
computing\cite{grandchallenge}. The results of those studies, in turn,
can be important to the understanding of detectable sources of
gravitational waves\cite{ligo}.

To the present date the only case that has been extensively studied is
the head-on collision, from rest, of two holes starting with the
initial value solution given by Misner\cite{misner}. The spacetime
growing out of those initial data has been computed by the techniques
of numerical relativity\cite{anninos_etal93}, and has been studied by
analytic
means\cite{price_pullin94}\cite{all}\cite{abrahams_cook94}.

The initial value solution of Brill and Lindquist\cite{bl} (hereafter
``BL''), like the Misner solution, represents two initially stationary
nonspinning holes.  Neither solution contains any initial radiation of
short wavelength compared to the characteristic size of the throats.
Outside the horizon the two initial value solutions can be thought of
as differing in the initial distortion of each throat caused by the
presence of the other throat.
There is in fact no solution of the initial value equations of general
relativity that is uniquely singled out as representing two initially
stationary holes. The Misner solution and BL solution are special
only in their mathematical convenience, and in the topological
properties of the geometry of the initial surface extended inside the
throats.  Specifically, the Misner solution may be thought of
as having a two-sheeted topology.  The two throats representing
the two black holes connect an upper ``physical'' sheet to
a single lower sheet isometric to the upper one.
In contrast, in the three-sheeted BL solution,
each of the throats connects from the
upper sheet to a separate lower sheet.  The isometry between
the two sheets in the Misner solution results from an infinite
series of image terms in the solution to the hamiltonian
constraint.  It is reasonable to expect that these terms
might lead to additional gravitational radiation, not present
in the BL solution.  Other physical consequences of the image
terms have been studied in Ref. \cite{ck}.

Here we extend the analytic study of collisions of holes to the case
of BL initial data.  There are two main justifications for doing this.
The first is that analytic answers are a useful aid to development of
the codes used in numerical relativity. The values reported here for
radiated energy can be tested against numerical codes for evolution of
axisymmetric initial data. For initially close black holes, it will be
interesting to see whether those codes agree with the analytic answers
as well as they do in the case of Misner initial data.

The second reason for some interest in the evolution of BL data is the
general question of the relationship of initial data to the generation
of gravitational radiation.  In astrophysically realistic problems the initial
data will necessarily come from some approximation scheme, such as
post-Newtonian solutions. Such an approach is justified if the
gravitational wave signal generated depends only on certain general
features of the initial data and is insensitive to many details
(e.g., topology). The
comparison of the evolution of BL and Misner data gives us a simple
model for studying this question, and an interestingly simple (though
limited) answer.

In the next section we describe the application of close-limit
perturbation theory to the evolution of BL initial data.  In Sec.~III
results are given for the radiation predicted by perturbation
theory. These results are compared with available fully numerical
results for the BL case, and are compared with analogous results
previously reported for collision from Misner initial data.

\section{Close-limit perturbation theory for BL initial data}
Like the Misner solution, the BL geometry is conformally flat and
takes the form $ds^2=\Phi^4\,ds_{\rm flt}^2$, where $ds_{\rm flt}^2$ is
the line element for flat 3-dimensional space, and where $\Phi$
satisfies the Laplace equation in the flat space. In terms of
spherical coordinates $R,\theta,\phi$, for $ds_{\rm flt}^2$, the
Misner or BL metrics can be written.
\begin{equation}\label{eq:conflat}
ds^2=\Phi^4(R, \theta;
\mu_0)\left(dR^2+R^2\left[d\theta^2
+\sin^2\theta\,d\phi^2
\right]\right)\ .
\end{equation}
For the BL geometry, the form of $\Phi$, aside from a factor of 2,
corresponds to the potential of Newtonian theory, with points of mass
$m$ at positions $z=\pm z_0$ on the $z$ axis:
\begin{equation}
\Phi_{\rm BL}=1+\frac{1}{2}
\left(\frac{m}{\sqrt{R^2\sin^2{\theta}+(R\cos{\theta}-z_0)^2}}
+\frac{m}{\sqrt{R^2\sin^2{\theta}+(R\cos{\theta}+z_0)^2}}\right)\ .
\end{equation}
For $R>z_0$ the square roots can be expanded in a power series in
$z_0/R$ and the BL 3-geometry written as
\begin{equation}\label{eq:BLinR}
ds^2_{\rm BL}=\left[1+\frac{M}{2R}\sum_{\ell=0,2,\dots}
\left(\frac{z_0}{R}\right)^{\ell}P_\ell(\cos{\theta})
\right]^4 \left(dR^2+R^2\left[d\theta^2
+\sin^2\theta\,d\phi^2
\right]\right)\ ,
\end{equation}
where the $P_\ell$ are the Legendre polynomials, and where $M\equiv2m$.

We next make a transformation of the radial coordinate $R$ to a new
coordinate $r$, as if we were transforming, in the Schwarzschild spacetime,
from isotropic coordinates to Schwarzschild coordinates:
\begin{equation}\label{eq:Rdef}
R=\left(\sqrt{r} +\sqrt{r-2M}\right)^2/4\ .
\end{equation}
It is convenient now to rewrite the  line element for the 3-geometry as
\begin{equation}\label{eq:BLds}
ds^2_{\rm BL}=\left[1+\frac{M/(2R)}{1+M/(2R)}\sum_{\ell=2,4,\dots}
\left(\frac{z_0}{M}\right)^\ell\left(\frac{M}{R}\right)^{\ell}P_\ell(\cos{\theta})
\right]^4 \left(\frac{dr^2}{1-2M/r}+r^2\left[d\theta^2
+\sin^2\theta\,d\phi^2 \right]\right)\ ,
\end{equation}
where the meaning of $R$ is given by (\ref{eq:Rdef}).

The geometry in (\ref{eq:BLds}) reduces to the Schwarzschild geometry
if the summation in the leading factor on the right is ignored. That
summation, then, contains the information about the deviations from
sphericity and is the starting point for close-limit nonspherical
perturbation calculations\cite{ap1}. In particular, the parameter
$\epsilon\equiv z_0/M$ can be considered an expansion parameter for
perturbation theory. If, for each multipole index $\ell$, we keep only
the leading order in $\epsilon$, the approximation to the BL initial geometry
takes the form
\begin{equation}\label{eq:BLlin}
ds^2_{\rm BL}\approx\left[1+\frac{2M/R}{1+M/(2R)}\sum_{\ell=2,4,\dots}
\left(\frac{z_0}{M}\right)^\ell\left(\frac{M}{R}\right)^{\ell}P_\ell(\cos{\theta})
\right] \left(\frac{dr^2}{1-2M/r}+r^2\left[d\theta^2
+\sin^2\theta\,d\phi^2 \right]\right)\ .
\end{equation}

In principle, for each multipole index $\ell$, one can read off the
metric perturbations (which are purely even parity) from
(\ref{eq:BLlin}), can construct Moncrief's\cite{moncrief74} gauge
invariant perturbation wave function $\psi_{\rm pert}$, and can evolve
that wave function with the Zerilli equation\cite{zerilli}. In
practice, this need not be explicitly carried out.  There is a
striking similarity between the expressions in
(\ref{eq:BLinR})-(\ref{eq:BLlin}) and the equivalent expressions for
the Misner geometry\cite{price_pullin94}\cite{all}. The single
difference is the coefficients in the series appearing in
(\ref{eq:BLinR})-(\ref{eq:BLlin}). For the Misner initial geometry
the coefficients are $\kappa_\ell(\mu_0)$. The dimensionless quantity $\mu_0$
parametrizes the initial  separation of the
throats, and the $\kappa$'s are functions given in Ref.\cite{all}.
The single change
\begin{equation}
(z_0/M)^\ell\rightarrow4\kappa_\ell(\mu_0)
\end{equation}
converts (\ref{eq:BLinR})-(\ref{eq:BLlin}) to their equivalent form
for the Misner case.  This means, for a given $\ell$, that $\psi_{\rm
pert}$ for the BL case has precisely the same form as for the Misner
case; the outgoing gravitational waves, according to perturbation
theory, are identical in shape. They differ only in a multiplicative
factor. Since power carried by outgoing waves is proportional to the
square of $\psi_{\rm pert}$ the results for BL infall, for each
$\ell$, can be found by multiplying the Misner results by
$[(z_0/M)^\ell/4\kappa_\ell(\mu_0)]^2$.
We note in passing that the ``forced linearization'' procedure
discussed in Ref. \cite{ap1} is, of course, also applicable to
the BL data.

This Misner-BL equivalence applies for any separation of the holes.
For large separations of the throats it is not surprising that the
gravitational waves generated by BL and by Misner initial data should
be similar. For small initial separations, however, there is a
significant difference between the 3-geometries of Misner and BL, and
it does seem strange that the gravitational waveforms should be
identical. Furthermore, it is for close initial separation that
perturbation theory is most applicable, so the prediction of identical
linearized waveforms is also a prediction about the actual waveforms.
How can such different initial conditions give rise to identical outgoing
waveforms?

It is important to realize that the linearized outgoing waves are
identical in form for each $\ell$, but the ratio of multipole
contributions differs for BL and Misner. In Fig.~1 this difference
in multipoles is shown quantitatively. For a given value of $\mu_0$
in the Misner geometry, an equivalent configuration for the BL
geometry is defined by setting the quadrupole amplitudes of $\psi_{\rm
pert}$ equal, i.e., by setting $(z_0/M)=2\sqrt{\kappa_2(\mu_0)}$. The
ratios of the BL amplitude to the Misner amplitude are then computed
for $\ell=4$ and $\ell=6$. (These amplitude ratios are in fact simply
$4\kappa_2(\mu_0)^2/\kappa_4(\mu_0)$ and
$16\kappa_2(\mu_0)^3/\kappa_4(\mu_0)$.)  At large separation the
amplitude ratios approach unity; this shows that in the limit of large
separation the external fields become identical in the two initial
geometries. For small separations, however, the BL solution has a
relatively smaller contribution due to higher multipole moments; its
geometry is more quadrupole dominated. Though this is a relatively
important difference in the initial geometry near the throats, it is
of little importance for the gravitational radiation. Even for the
Misner initial conditions, the radiation is heavily quadrupole
dominated. It is possible that the lesson of this example has
a broader generality: the outgoing radiation can be insensitive to
many details of the initial data and even for strong field sources a
knowledge of the quadrupole moment may be all that is needed.

It is worth asking whether there is any deep physical meaning in the
fact that the only difference between the BL and Misner linear
perturbations is the ratio of the multipole amplitudes. This follows
from the fact that for a conformally flat 3-metric, with the form
(\ref{eq:conflat}), the factor $\Phi$ satisfies the flat space
Laplacian. If the solution is axisymmetric and asymptotically flat it
must be of the form
$\sum(\alpha_\ell/R^{\ell+1})P_\ell(\cos{\theta})$; solutions can
differ only in the values of the constants $\alpha_\ell$.  So the
striking similarity of the BL and the Misner perturbations is a direct
result of the choice of the conformally flat form
(\ref{eq:conflat}). This choice is dictated by convenience, and need
not be made in principle. For more general momentarily stationary
initial geometries the linearized waveforms for each multipole will
have different appearance.  For example, one could generate
valid initial data representing a Schwarzschild spacetime
with a nonconformally-flat perturbation by choosing an arbitrary
(small) metric perturbation and solving the linearized hamiltonian constraint
for the conformally-flat part of the perturbation. The gauge-invariant
function would then be computed from the full perturbation.

\section{Radiation Energy: BL vs. Misner}
The first, and most difficult, step in comparing radiation from the
two initial value sets is to decide on the basis for comparison: How
does one compare a BL problem with a particular value of $z_0/M$ with
a Misner problem of a particular $\mu_0$? At large separations it is
not difficult; one can compare BL and Misner configurations in which
the masses and separation of the holes are identical. For small
separations, however, the separation of the holes is somewhat
ambiguous.  To deal with small, as well as large, separations we
choose a reasonably natural and convenient specific measure of the
separation $L$: the proper distance along the symmetry axis, between
the outermost disjoint marginally outer-trapped surfaces
around each throat. (For $z_0/M$ less than about 0.4, a single
apparent horizon encompasses both holes).
The locations of the marginally outer-trapped surfaces was
found using a standard shooting technique applicable to
axisymmetric spatial slices\cite{ah}.
We characterize both BL
and Misner configurations with $L/M$, where $M$ is the mass of the
spacetime. It is, of course, interesting not only to compare the
linearized predictions for BL against those for Misner, but also to
compare both against the results of numerical solutions of the fully
nonlinear field equations. For the Misner initial geometry the
numerical results are known from the work reported in
Ref.~\cite{anninos_etal93}. For BL initial conditions two data points
are available: cases c2 and c4 of from Ref.\cite{ast95}.
These numerically generated spacetimes have euclidean spatial topology,
with initial data consisting of spherical (in the conformal space)
collisionless matter configurations.  When the initial configurations are
sufficiently compact, the matter is all inside disjoint apparent
horizons  and the external 3-geometry is identical to the BL data.

For clarity, the results are presented in three separate
figures. Figure 2 shows the comparison of perturbation results and
numerical results for the Misner case.  The perturbation energies
($E/M\approx0.0251\kappa_2^2(\mu_0))$ are those of
Ref.~\cite{price_pullin94}, except that energy has been plotted as a
function of $L/M$, rather than of $\mu_0$. The numerical data are
those of Refs.~\cite{anninos_etal93} and \cite{all}.  Figure 3 shows
the analogous results for the BL case, for which the energy is
$E/M\approx0.0251[(z_0/M)^2/4]^2$.  The two ``numerical'' data points
here are those of Ref.~\cite{ast95}.

Figure 2 shows that for the Misner case, linearized predictions begin
to diverge from the fully numerical results at around $L/M=4$. It is
fortunate that the numerical results available for the BL case are for
$L/M$ in the range 3--4. From Fig.~3 we can infer that for $L/M$ less
than around 3, the agreement between linearized and numerical results
is very good for BL collisions, and for $L/M$ above 4 there is
significant disagreement. In this sense there is little difference
between Misner and BL cases.  Figure 4 shows the perturbation theory
comparison of Misner and BL cases. This figure shows that there is
little difference between the predicted radiation when $L/M$ is
greater than around 2. It is, therefore, not surprising that the
agreement between numerical and perturbation results, which breaks
down well above $L/M=2$, does not distinguish between BL and Misner
collisions. It is also not surprising that in BL collisions, as in
Misner collisions\cite{all}, the radiation is always quadrupole
dominated.  (The large values of hexdecapole energy in Fig.~3 occur only
at separations large enough that linearized theory wildly
overestimates radiation.)

The results in Fig.~4 would seem to suggest that, for black holes
initially close, BL initial conditions lead to less radiation than
Misner black holes, as expected by the presence of image
terms in the Misner solution. An alternative interpretation
is that, for equal
radiation, the initial separation of the apparent horizons is greater
in the BL case than in the Misner case. Since equal radiation implies
equal quadrupole moments, this means that the different multipole
structure of the BL and Misner geometries makes the proper distance
between apparent horizons larger in the BL case when quadrupole
moments are equal. In this sense then, Fig.~4 is more of a depiction of
proper distances than of radiation.

This motivates asking whether there is a way of comparing BL and
Misner scenarios that is better, or at least different, from using
$L/M$. Another physically meaningful measure of how close the initial
throats are is the gravitational binding energy.
The gravitational binding energy is the difference between the
ADM energy of an initial data set representing two
black holes at finite separation and the energy of an
initial data set with the holes infinitely separated
(the sum of the bare masses of the holes).
For BL data this is given by\cite{bl}
\begin{equation}
{E_B \over M} =  -{1 \over 8 z_0}.
\end{equation}
For Misner data one has\cite{lind63}
\begin{equation}
{E_B \over M} = - {\sum_{n=1}^{\infty} (n-1){\mathop{\rm
                   csch}\nolimits}{n \mu_0} \over \sum_{n=1}^{\infty}
                   {\mathop{\rm csch}\nolimits}{n \mu_0} }
\end{equation}

Radiated energy is plotted against
binding energy in Fig.~5, but the results give a picture very much
like that of Fig.~4. In particular, for small initial separations
(tightly bound initial configurations) there is less energy radiated
from a BL collision than from a Misner collision. For large initial
separations (small binding energies) the difference in radiated energy
is small for configurations with the same binding energy. The BL and
Misner cases become significantly different (say by a factor of 2) for
binding energy (binding energy/$M\approx-1.5$) that corresponds
roughly to the point ($L/M\approx1.3$) at which the BL and Misner
energies separate in Fig.~4.

We thank Greg Cook, Jorge Pullin, Ed Seidel, Stuart Shapiro,
Wai-Mo Suen, and Saul Teukolsky for helpful
discussions.  AMA was supported by NSF grant PHY 93-18152/ASC 93-18152
(ARPA supplemented), and RHP was supported by NSF grant PHY95-07719.

%%%%%%%%%%%%%%%%%%%%%%

%%%%%%%% FIGURES
\begin{figure}
\caption{Ratio of amplitudes of $\psi_{\rm pert}$ for BL and Misner
geometries. For equal amplitudes of $\ell=2$, amplitude ratios are shown
for $\ell=4$ and $\ell=6$. }%%%FIGURE 1
\end{figure}

\begin{figure}%%%%%%%%FIGURE 2
\caption{Gravitational radiation energy emitted during the head-on
collision of two black holes starting from the Misner initial
conditions. Results are shown for close-limit perturbation theory
(continuous curve) and for numerical relativity (isolated points).}
\end{figure}

\begin{figure}%%%%%%%FIGURE 3
\caption{Gravitational radiation emitted during the head-on collision
of two black holes starting from BL initial conditions. Results are
shown for close-limit perturbation theory (continuous curve), and two
values are shown from numerical relativity.}
\end{figure}

\begin{figure}%%%%%%%FIGURE 4
\caption{Comparison of perturbation theory predictions for radiated
energy from Misner and from BL initial conditions. Results are given for both
$\ell=2$ and 4 multipoles.}
\end{figure}

\begin{figure}%%%%%%%FIGURE 5
\caption{Gravitational radiation emitted from Misner and BL initial
conditions are plotted as a function of the binding energy of the
initial configuration divided by the mass of the spacetime.}
\end{figure}

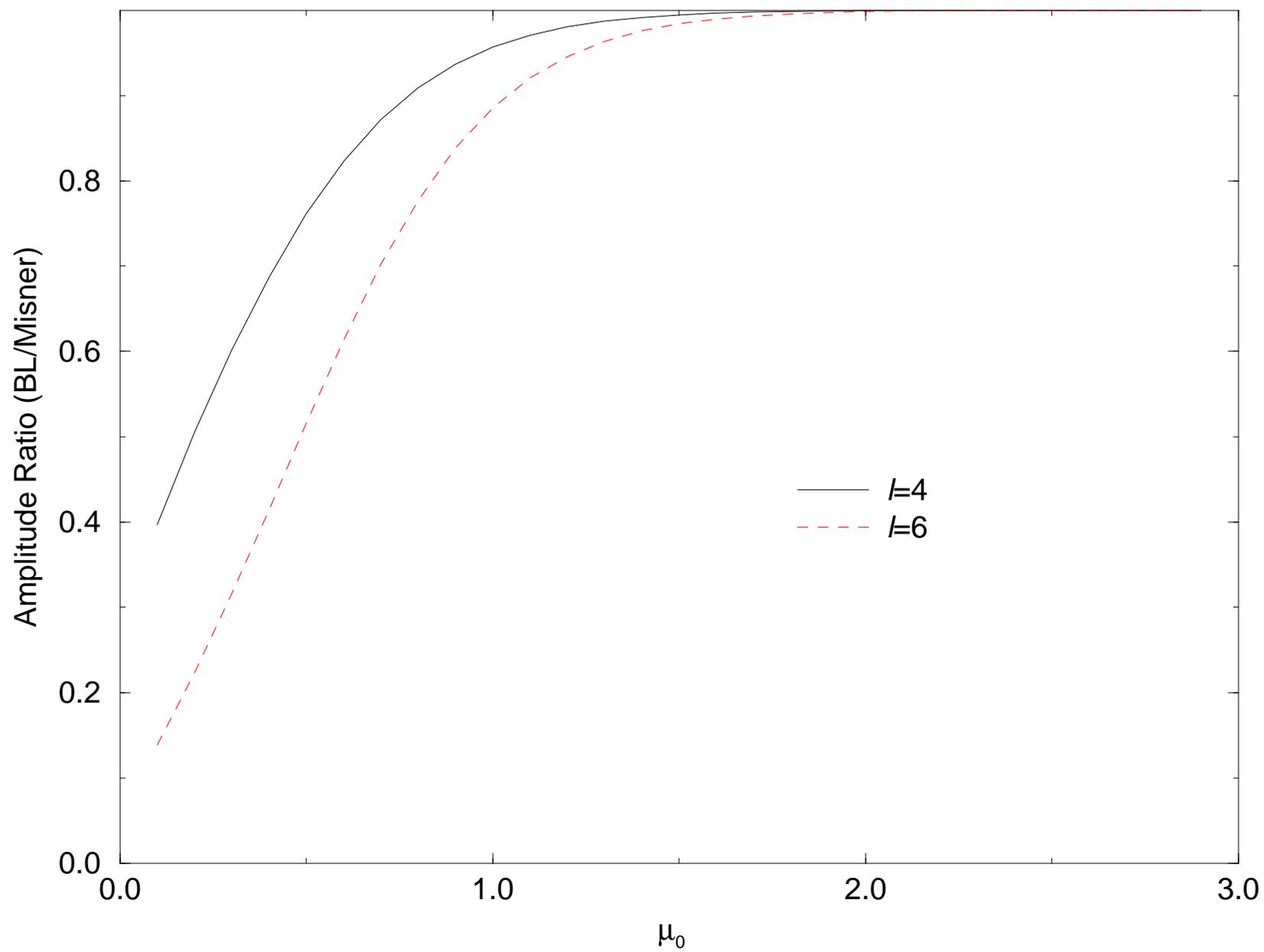
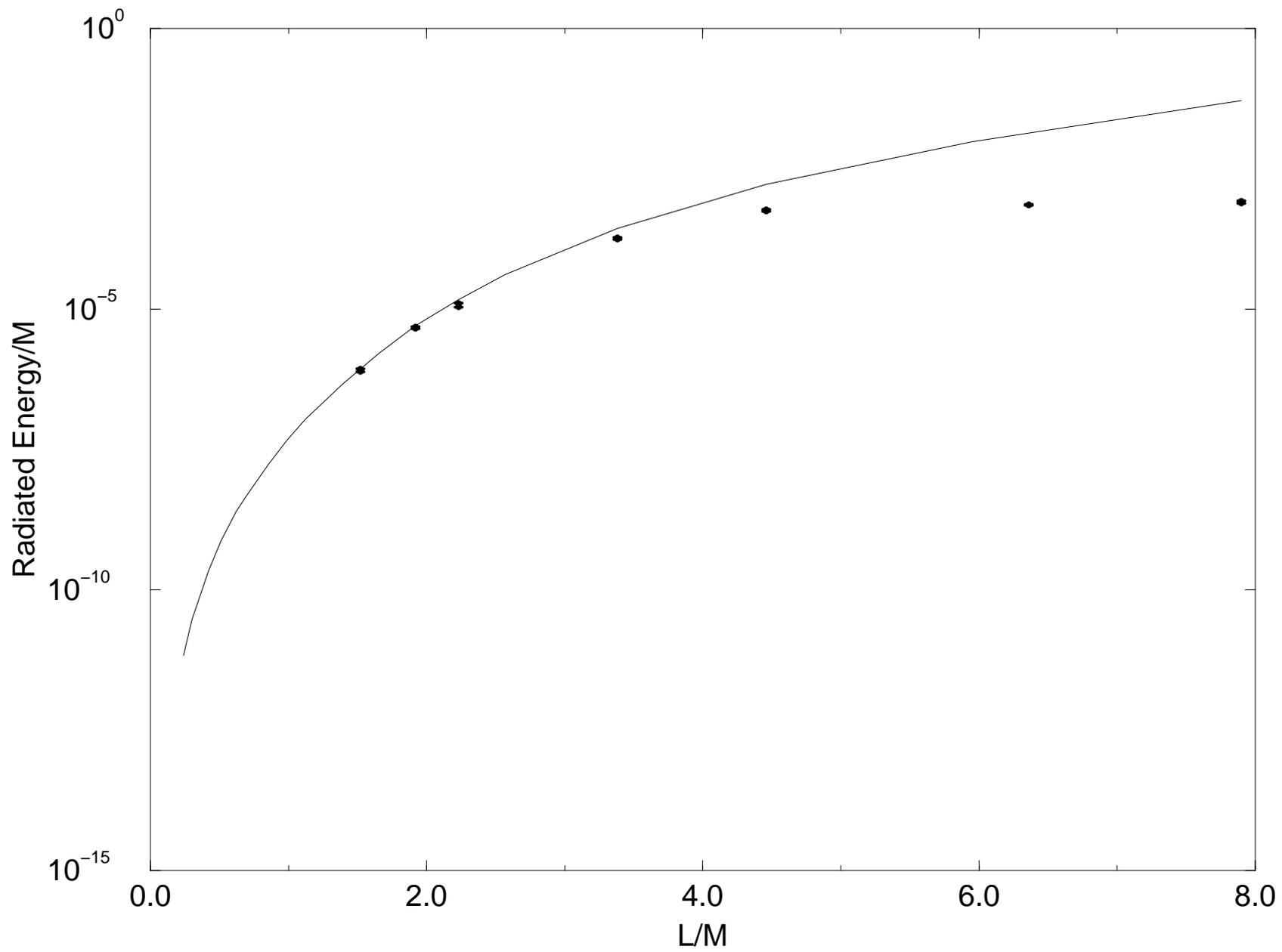
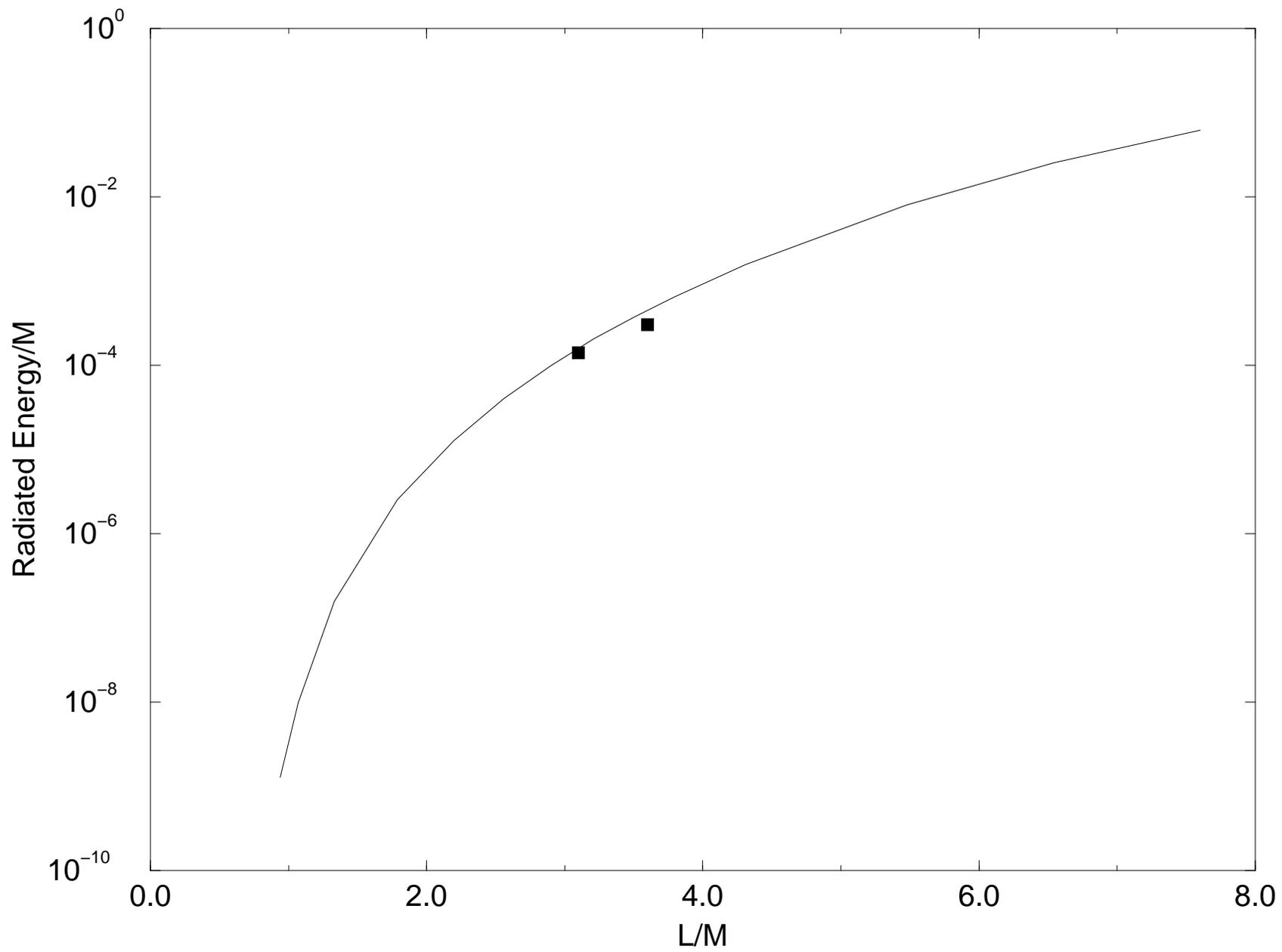
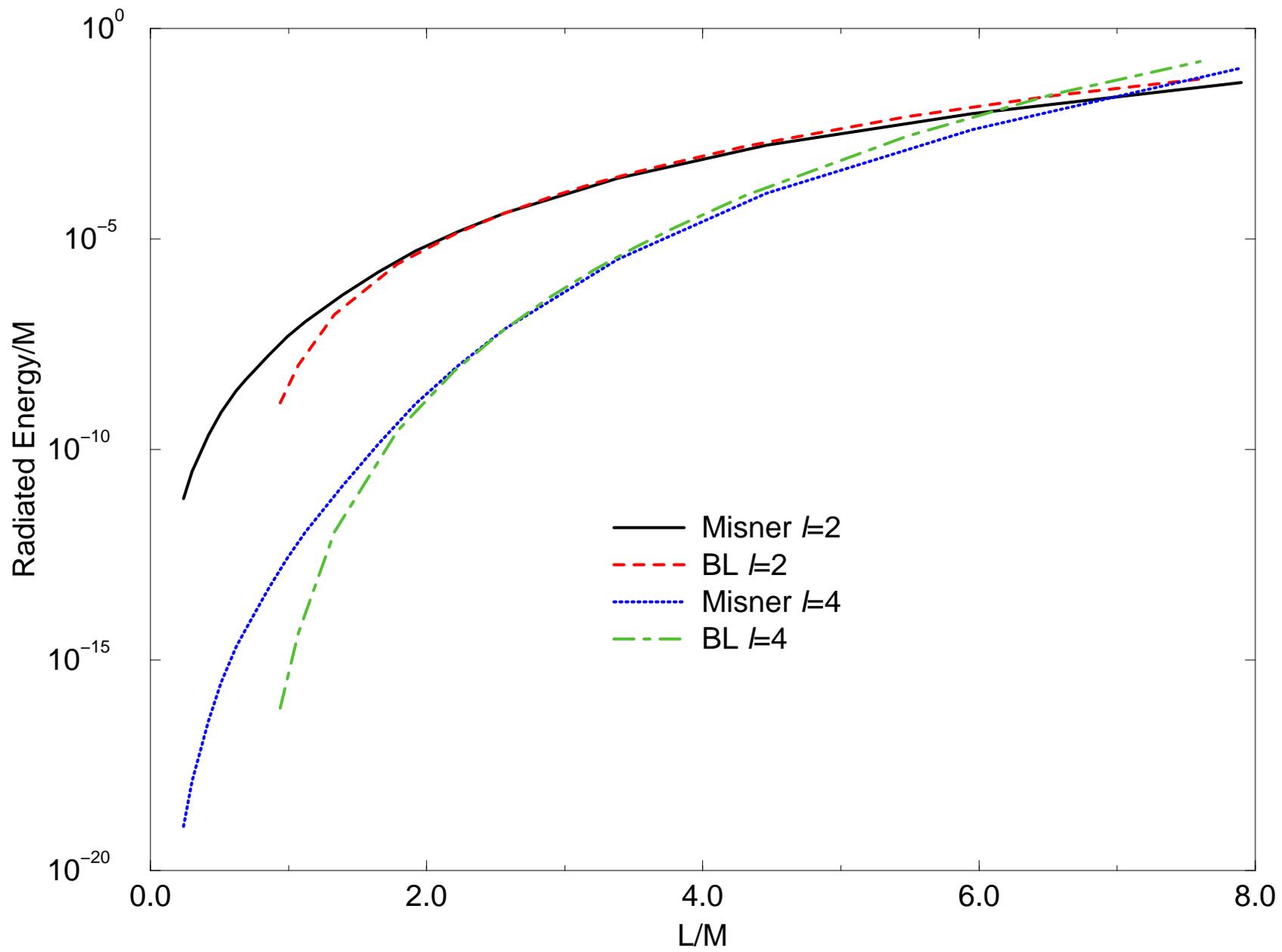
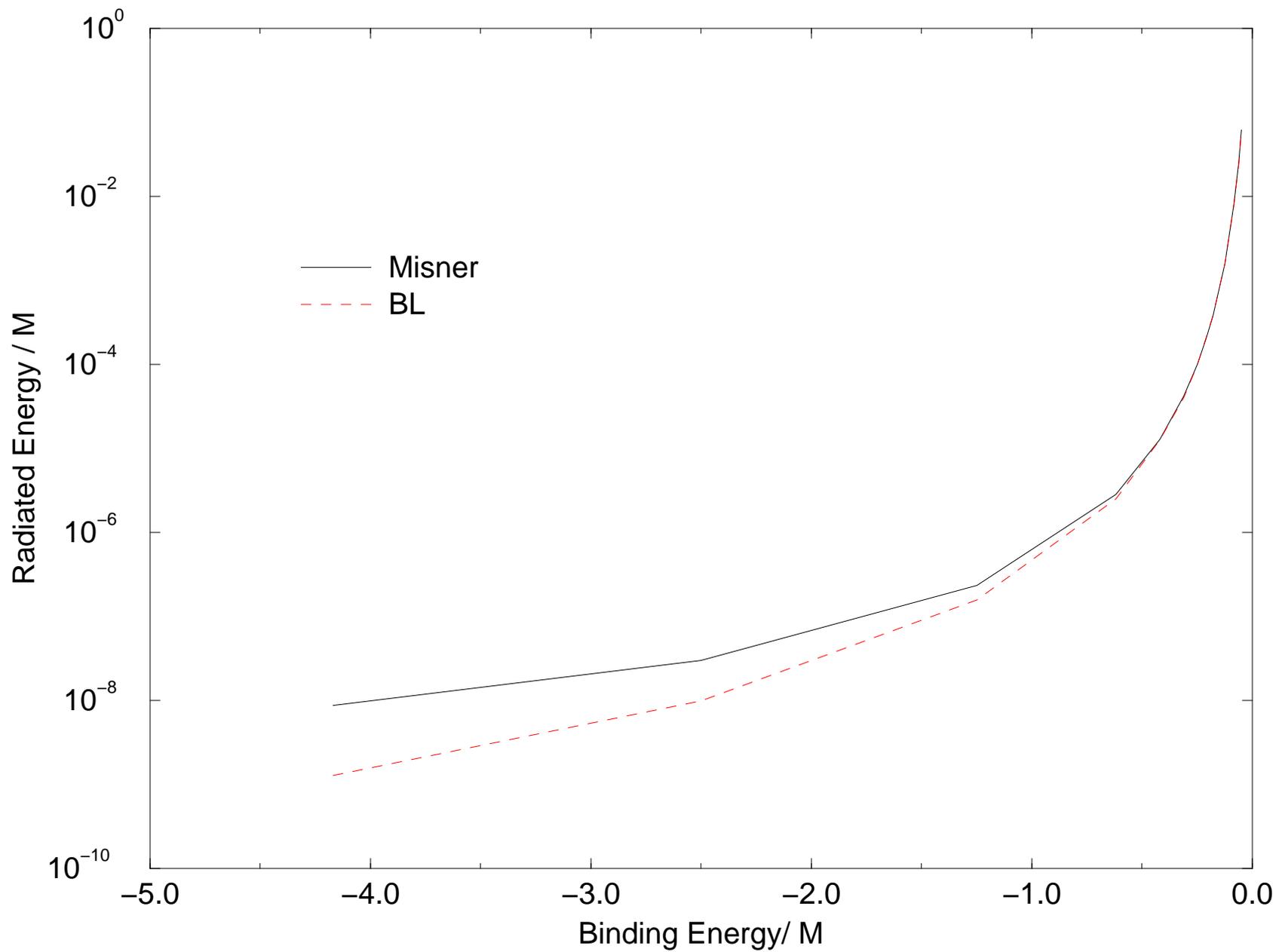
\end{document}